# Is Lithium Sulfide a MgB$_2$-like Superconductor?

O. P. Isikaku-Ironkwe[1, 2]
[1]The Center for Superconductivity Technologies (TCST)
Department of Physics,
Michael Okpara University of Agriculture, Umudike (MOUAU),
Umuahia, Abia State, Nigeria
and
[2]RTS Technologies, San Diego, CA 92122

## Abstract

Lithium Sulfide, Li$_2$S, is an anti-fluorite semiconductor with a band-gap of 3.865 eV. It also has exactly the same valence electron count, Ne, and atomic number, Z, as magnesium diboride, MgB$_2$. Both have almost the same formula weight. This qualifies Li$_2$S as a magnesium-diboride like material. Li$_2$S passes the same computational material specific test for superconductivity as MgB$_2$. Using our recently developed symmetry rules for searching for superconductors, we predict that Li$_2$S, with electronegativity of 1.5, will be superconducting with a Tc of 33.8K, if double gapped or 16.9K if single gapped.

## Introduction

The discovery of MgB$_2$ as a superconductor [1], long overlooked by experts [2], led to an avalanche of renewed interest and publications [3] in binary and ternary superconductors in a relatively short time. Though the electronic structure and properties were quickly deciphered, the progress of reverse engineering of MgB$_2$-like superconductors was not equally a successful venture [4, 5, 6]. In the unsuccessful search, structure and iso-valency were the primary theoretical model and experimental strategy [5 - 8]. Using a new model based on similarities of electronegativity, valence electrons, atomic number and formula weight [9, 10] we have been able to predict many MgB$_2$-like superconductors awaiting experimental verification. Two of them are Lithium magnesium nitride LiMgN [11] and LiBSi [12]. In this paper, we review the structure and properties of another material, lithium sulfide Li$_2$S. We then apply the material specific characterization analysis scheme [10] and compare Li$_2$S, with MgB$_2$. We apply the symmetry rules for similar superconductors [10, 11], and estimate the Tc of Li$_2$S.

## Structure and Properties of Li$_2$S

Lithium sulfide is a much studied material [13, 14], though never tested for superconductivity. Li$_2$S can exist in two forms: orthorhombic and cubic. The orthorhombic form [15] belongs to space group Pmnb and has dimensions: a = 3.808Å; b = 6.311Å; c = 7.262Å. It has density of 1.75g/cm$^3$. The cubic version [16] has density of 1.63g/ cm$^3$, belongs to space group Fm-3m and has cubic dimensions 4.046Å. The electronic structure and density of states [16] indicate that cubic Li$_2$S is an indirect band-gap semiconductor with a band gap of 3.865 eV. Lithium sulfide melts between 900 – 975 degrees centigrade.

## Symmetry Rules and Tc Estimation

Guided by atomic number symmetry we find that Li$_2$S has same average atomic number as MgB$_2$. A MSCD [10] characterization of Li$_2$S yields the data displayed in table 1 and figure 1, showing that it is a MgB$_2$-like material. Earlier [10], we derived a formula for estimating maximum Tc, given by

$$T_c = x \frac{Ne}{\sqrt{Z}} K_o \qquad (1)$$

where $x$, Ne, Z are averages respectively of electronegativity, valence electron count and atomic number and $K_o$ is a parameter that determines the value to Tc. Ko = n(Fw/Z) and n is dependent on the family of superconductors. Fw represents formula weight of the superconductor. For MgB$_2$, Ko = 22.85 and Fw/Z is 6.26, making n = 3.65. The computational material specific test for superconductivity is that $0.75 \frac{Ne}{\sqrt{Z}} < 1.0$. Lithium sulfide meets this test range, implying that it is a superconductor. The applicable symmetry rule [10, 11] is:

*"If two or more materials have the same average valence electrons Ne, and atomic number Z, then their Tcs will be proportional to their electronegativities."* Applying this rule, we find that the Tc of Li$_2$S will then be approximately 1.5/1.7333 of 39K or 33.8K.

## Discussion

Sulphur has been found under pressure to show superconductivity up to 17K [17]. Lithium too is known to be a metallic superconductor [18] with Tc of 20K under pressure of 48 GP. One would expect that a combination of lithium and sulphur should produce an even higher Tc. It is interesting to observe that the band-gap of $Li_2S$ of 3.865 eV is close to that of LiMgN [11] which we predicted as having same Tc as $MgB_2$. Generally we have found that ternary or three-atom materials with whose atomic numbers add up to 22 tend to have their valence electrons adding up to 8, just like magnesium diboride. Matthias [19] who was one of the first to study superconducting sulfides, used the concept of symmetry of crystals in the Periodic Table to raise Tc in a number of sulfides. Lithium and its compounds find strong commercial applications in the battery industry. If confirmed, this will be the first lithium-sulphur compound superconductor.

## Conclusion

Simple symmetry rules [10, 11], involving electronegativity, valence electrons and atomic number and formula weight, discovered to apply to many other superconductors, strongly suggest that Lithium sulfide, $Li_2S$, though an ionic compound may prove to be a superconductor with Tc of 33.8K, comparable to that of magnesium diboride.

## Acknowledgements

Discussions with A.O.E. Animalu on symmetry in mathematics, physics, chemistry and nature influenced my search strategy. My thanks go to J.B. O'Brien at Quantum Design who provided literature support and useful discussions while M.J. Schaffer, then at General Atomics, fully sponsored this research.

| Material | $\chi$ | Ne | Z | Ne/$\sqrt{Z}$ | Fw | Fw/Z | Tc(K) | Ko |
|---|---|---|---|---|---|---|---|---|
| 1 MgB$_2$ | 1.7333 | 2.667 | 7.3333 | 0.9847 | 45.93 | 6.263 | 39 | 22.85 |
| 2 Li$_2$S | 1.5 | 2.6667 | 7.3333 | 0.9847 | 45.95 | 6.266 | 33.8 | 22.85 |

Table 1: Material specific characterization datasets (MSCDs) for MgB$_2$ and Li$_2$S. Note that Ne and Z are the same for MgB$_2$ and Li$_2$S. Their formula weights (Fw) differ by just .02. The Tc of Li$_2$S was computed with equation (1) and the symmetry rule. The Ne/$\sqrt{Z}$ tells us that Li$_2$S is a superconductor like MgB$_2$.

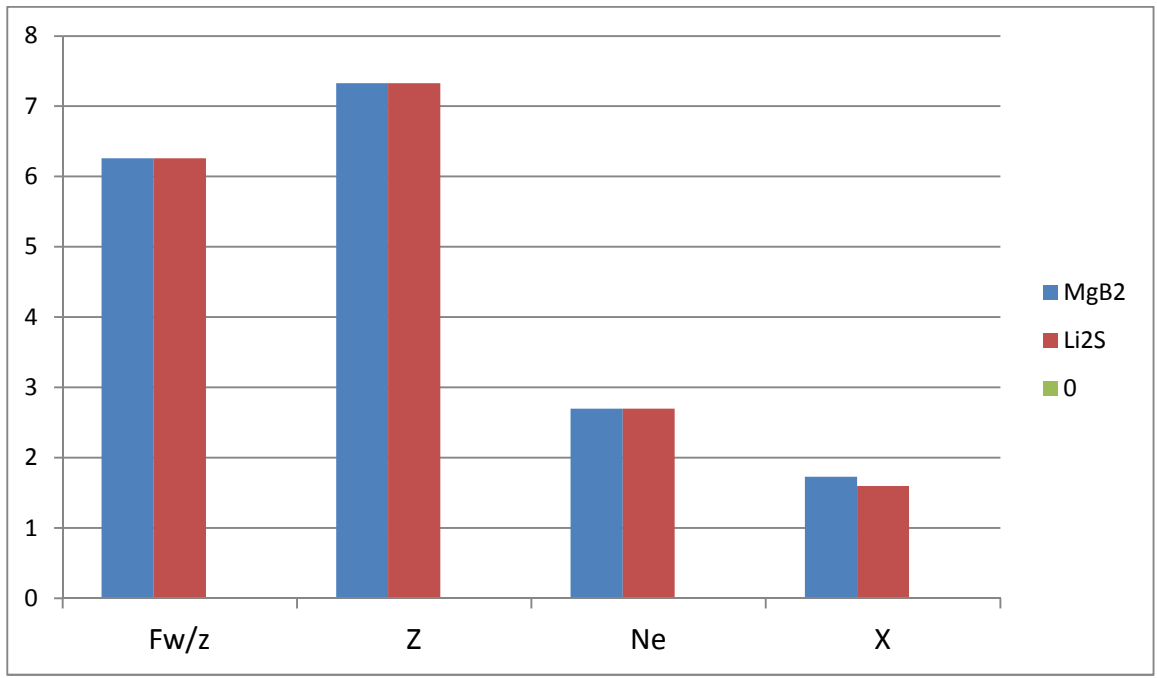

Figure 1: MgB$_2$ and Li$_2$S compared in terms of Fw/Z, Z, Ne and X. This similarity suggests that their Tcs will be close.